State and group dynamics of world stock market by principal component analysis

Ashadun Nobi[1,2] and Jae Woo Lee[1,a)]

[1]*Department of Physics, Inha University, Incheon 402-751 South Korea*

[2]*Department of Computer Science and Telecommunication Engineering, Noakhali Science and Technology University, Sonapur Noakhali-3802, Bangladesh*

Abstract

We study the dynamic interactions and structural changes in global financial indices in the years 1998-2012. We apply a principal component analysis (PCA) to cross-correlation coefficients of the stock indices. We calculate the correlations between principal components (PCs) and each asset, known as PC coefficients. A change in market state is identified as a change in the first PC coefficients. Some indices do not show significant change of PCs in market state during crises. The indices exposed to the invested capitals in the stock markets are at the minimum level of risk. Using the first two PC coefficients, we identify indices that are similar and more strongly correlated than the others. We observe that the European indices form a robust group over the observation period. The dynamics of the individual indices within the group increase in similarity with time, and the dynamics of indices are more similar during the crises. Furthermore, the group formation of indices changes position in two-dimensional spaces due to crises. Finally, after a financial crisis, the difference of PCs between the European and American indices narrows.

The dynamic changes of the world stock indices have been studied by the analysis of the cross-correlations among the times series of the stock indices. In this paper, we apply the principal component analysis (PCA) to cross-correlation coefficients of the world stock indices. We observe the variances of the PCs over the time. The first PC shows a drastic change during the global financial crisis period. We identify the transition of market state by the sharp change in PC coefficient. We observe the frequent change of PC coefficient in the American and Asian indices. However, the European indices remain the stable state over time. Using the first and second PC coefficient, we observe three regional groups whose indices are closely related. The PCA is a good tool to identify market state and to determine the subsets of global stock indices.

I. Introduction

Scientists from many fields have been attempting to understand the dynamics of financial markets for the past two decades [1-7]. There are many reasons for wanting to understand correlations in price behaviors. Due to estimated financial risk, statistical dependencies between stocks are of particular interest. Statistical dependencies within the market change with time due to the non-stationary behavior of the markets, which complicates the analysis. As a result, different kinds of methods and techniques are applied to analyze financial systems and extract its

a) Email: jaewlee@inha.ac.kr

contained information [7-12]. Principal component analysis (PCA) is one of the established methods to characterize the evolving correlation structures of financial markets and to measure the associated systemic risk [13-16]. Generally, PCA is a multivariate statistical technique particularly useful for analyzing the patterns of complex and multidimensional relationships that transform a large number of related observable variables into a smaller set of observable composite dimensions that can be used to represent their interrelationships. This method is used in different branches of science such as engineering, chemistry, and food technology to reduce the large dimensionality of the data sets and to characterize systems. In finance, most studies are carried out on financial sectors such as banks, brokers, insurance companies or hedge funds in order to measure systemic risks, arbitrage pricing theory and portfolio theory [17-19]. A recent study applied the PCA method to 10 different Dow Jones economic sector indices and showed that the larger was the change in PC1, the greater was the increase in systemic risk [16]. Here, we apply the PCA method to global financial indices to identify the market state of each index and to classify the groups of indices. The returns of some markets are particularly associated with groups of nations, and the returns of each one are based on the returns of the associated group. The application of the PCA method to global financial indices has been successfully performed [14, 20]. In reference 14, the authors estimate the global factor as the first component using principal component analysis. In reference 20, the authors use a diverse range of asset classes such as equity indices, bonds, commodities, metals, currencies, etc. and consider weekly time series. They investigate the correlations using random matrix theory and the PCA method and identify the notable changes in assets during the credit and liquidity crisis in 2007-2008. In another recent study, the PCA method is applied to financial indicators of Europe, Japan and the United States, and the group of indicators is identified [21]. In our work, we use daily closing prices of 25 global indices from 1998-2012, analyzed in one-year time windows, and identify significant changes in market state due to external or internal crises over the entire period. Generally, different kinds of network techniques such as the minimum spanning tree, hierarchical method, planar maximal graph, and threshold method are applied to segment global equity markets [22, 23]. For segmentation of global indices, we use the components of the first two PCs, a unique approach for complex non-stationary systems.

The rest of the paper is organized as follows: the financial data are discussed in Section II. The method of PCA is explained in Section III. The market state is analyzed in Section IV. The group dynamics are discussed in Section V. Finally, we draw our conclusions in the final section.

II. DATA ANALYSIS

We analyze the daily closing prices of 25 global indices from January 2, 1998 to December 20, 2012. These global financial indices are as follows: 1. Argentina (ARG), 2. Austria (AUT), 3. Australia (AUS), 4. Brazil (BRA), 5. Germany (GER), 6. India (IND), 7. Indonesia (INDO), 8. Israel (ISR), 9. South Korea (SKOR), 10. Malaysia (MAL), 11. Mexico (MEX), 12. The Netherlands (NETH), 13. Norway (NOR), 14. United Kingdom (UK), 15. United States (US), 16. Belgium (BEL), 17. Canada (CAN), 18. China (CHN), 19. France (FRA), 20. Hong Kong (HKG), 21. Japan (JPN), 22. Singapore, 23. Spain (ESPN), 24. Switzerland (SWZ), and 25. Taiwan (TWN). The sequence numbers $(1, \cdots, 25)$ and the abbreviations $(ARG, \cdots, TWN)$ of the indices above are used to label the indices in the pictures. The data are collected from Ref. 24. To design an equal-time cross-correlation matrix, we exclude public holidays for which 30% of the markets are closed. Again, we add some days for a specific market if that market is closed on a particular day. In such cases, we consider the final closing prices for that day. Thus, we consider all indices on the same date and filter the data as in Ref. 25. We examine daily returns for the indices, each containing approximately 260 records for each year.

## III. Principal Component Analysis in the Stock Market

We analyze the daily logarithmic return, which is defined for index $i$ as

$$r_i(t) = \ln[I_i(t)] - \ln[I_i(t-1)], \tag{1}$$

where $I_i(t)$ is the closing price of index $i$ on day $t$. The normalized return for index $i$ is defined as

$$r'_i(t) = (r_i(t) - <r_i>)/\sigma_i, \tag{2}$$

where $\sigma_i$ is the standard deviation of the stock index time series $i$ over the time window, and the symbol $<\cdots>$ denotes average over the time window. Then, the normalized returns matrix $R$ is constructed from the time series $r'_i$ with the dimensions $N \times T$. This variable can be used to calculate the correlation matrix as

$$C = \frac{1}{T} R R^T, \tag{3}$$

where $R^T$ is the transpose of $R$. We next diagonalize the $N \times N$ correlation matrix $C$ in the form

$$C = VMV^T, \tag{4}$$

where $M$ is the diagonal matrix of eigenvalues $\lambda_i = (\lambda_1, \lambda_2, \cdots, \lambda_N)$ in descending order, and $V$ is an orthogonal matrix of the corresponding eigenvectors. Each eigenvalue and the corresponding eigenvector can be written as

$$\lambda_i = v_i C v_i^T = v_i Cov(R_t) v_i^T = Var(v_i^T R_t) = Var(y_{k,t}), \tag{5}$$

where $y_{k,t} = v_i^T R_t$ is the $k_{th}$ principal component [26]. The eigenvalue $\lambda_i = Var(r'_{i,t})$ indicates the portion of the total variance of $R_t$ contributing to the principal component $y_{i,t}$. Then, the total variance of the returns R for N assets is

$$\sum_{i=1}^{N} Var(r'_i) = tr(C) = \sum_{i=1}^{N} \lambda_i = N. \tag{6}$$

The proportion of total variance in R explained by the $k^{th}$ PC is $\lambda_i/N$. Figure 1 shows the fraction of variances explained by the first three PCs as a function of time. The variances explained by the first PC range from 30% to 60%, and that for the second PC is around 10%. The contribution of variances for the next PCs gradually decreases. The variance explained by the first PC shows a high value during the Russian crisis in 1998. After that, the variances show an increasing trend until 2004. A significant change in variance is observed in 2006 when the mortgage crisis originates. The rest of the year's variances explained for the first PC show high values due to the mortgage crisis, global financial crisis, and European sovereign debt crisis. The variance peaks are observed during the global financial crisis (2008) and European sovereign debt crisis (2011). The variance explained by the first PC increases around 36% from 2005-2008. Over the same period, the variances explained by the second PC increase by 26%. However, the variances for the third PC do not show any significant variation due to crises.

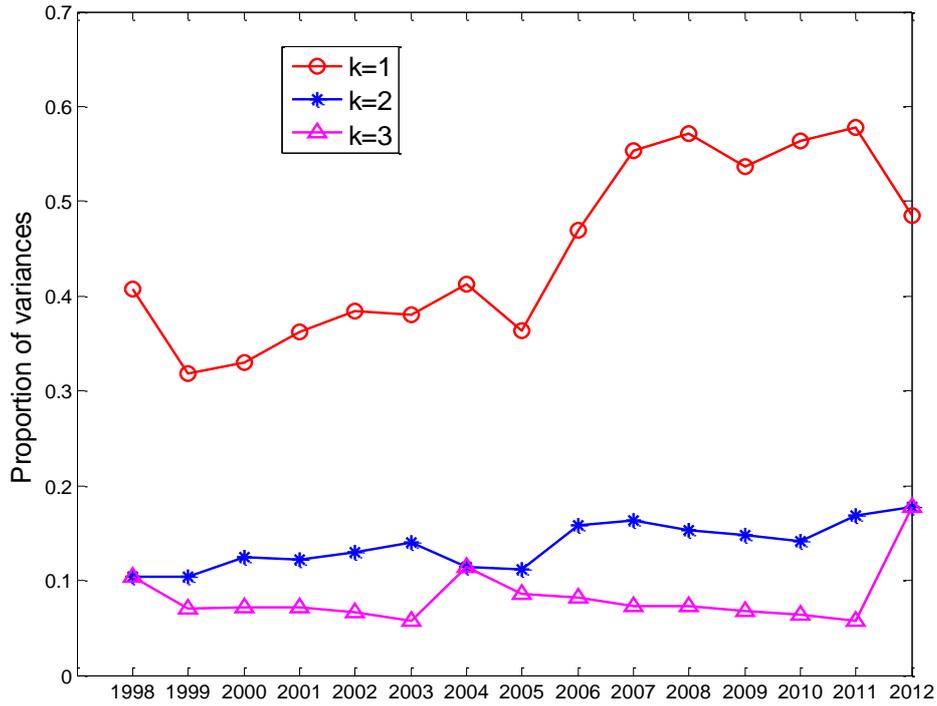

FIG. 1. The fraction of the variance in return explained by the first, second and third PCs. Significant changes in the variance in return are observed in 2006, when the mortgage crisis originates.

IV. PC Coefficient and State of the Market

The correlations $C_1(r_i', y_k)$ between indices and PCs closely related to PC coefficients imply the contribution of each index to each PC. These correlations are known as the weight or load of each index on the PCs [20]. The covariance matrix of the return R with PCs Y can be written as

$$Cov(Y, R) = \frac{1}{T} Y R^T = \frac{1}{T} V R R^T = VC = VV^T MV = MV. \qquad (7)$$

Thus, the cross correlation of the returns of asset $i$ and the $k_{th}$ PC can be written as $Corr(r_i', y_{1,t}) = \sqrt{\lambda_i}\, v_{i1}$, , which is equal to the PC coefficients scaled by the appropriate eigenvalue. To identify the market state, we consider the correlations between indices and PCs for the largest eigenvalues. Because correlations are an indicator of the market state, the temporal change in such correlations or PC coefficients can identify the states of the market for each index. Since the first eigenvalues indicate the market dynamics, the largest PCs are appropriate to determine the market state over time. We define the change in the correlations of each index contributing to PC1 between two time periods as $\Delta C_1(t_1, t_2) = C_1(t_2) - C_1(t_1)$. The gray-scale configurations of the change in correlations are presented in Fig. 2. This new representation gives a complete overview of the change in correlations of global financial indices over the 15 year period. It allows us to compare the state of the market at different times. The change in sign of $\Delta C_1$ implies that the market oscillates with time. The high values of $\Delta C_1$ indicate a significant change in market state. To understand the market state of a country using the figure, pick an index on the vertical axis of the gray-scale configuration and move in a horizontal line. Light yellow or light green areas indicate no significant change in the market state, whereas dark blue or dark red areas denote a sharp change in the market state. We further identify the financial crises with dark or light red areas. The recovering states are represented by dark or light blue areas. Our gray-scale configuration shows that some countries (India, Indonesia, Malaysia, Austria, and Argentina) frequently change market state, while some countries (Germany, UK, France, Hong Kong, The Netherlands, Singapore) do not show a sharp change in market state over the observation period. On the other hand, although there is no associated global crisis, the Indonesian and Malaysian markets experience a change in market state from 2003-2005. These changes may be due to internal crises. The indices that do not change market state due to external or internal information can be regarded as risk-free indices and are selected for portfolio investment. A Russian crisis occurred in 1998, followed in 1999 by a whole-market transition, which implies market recovery. Next, the dot-com bubble originated in 2000, and some markets enter into crisis states (South Korea and India). However, in the same period, some European markets show a recovery state after the Russian crisis (Austria, Belgium, and Switzerland). The market shows crisis and recovery states from 2000 to 2005 due to the effect of the dot-com bubble or internal crises, as shown in Fig. 2. A significant change in the states of the markets is observed starting in 2006 when the mortgage crisis originated. During the mortgage crisis of 2007, most of the markets enter into a crisis state. During the global financial crisis, most of the markets remain unchanged except those of Indonesia, Malaysia, Singapore and Taiwan in 2007. Although the market shows a recovering trend in 2009, there are still a few markets experiencing a crisis state. Due to the European sovereign debt crisis that originated in 2010 in some states of Europe, most of the markets are in a crisis state until 2011. Again, the markets show a recovering trend in 2012.

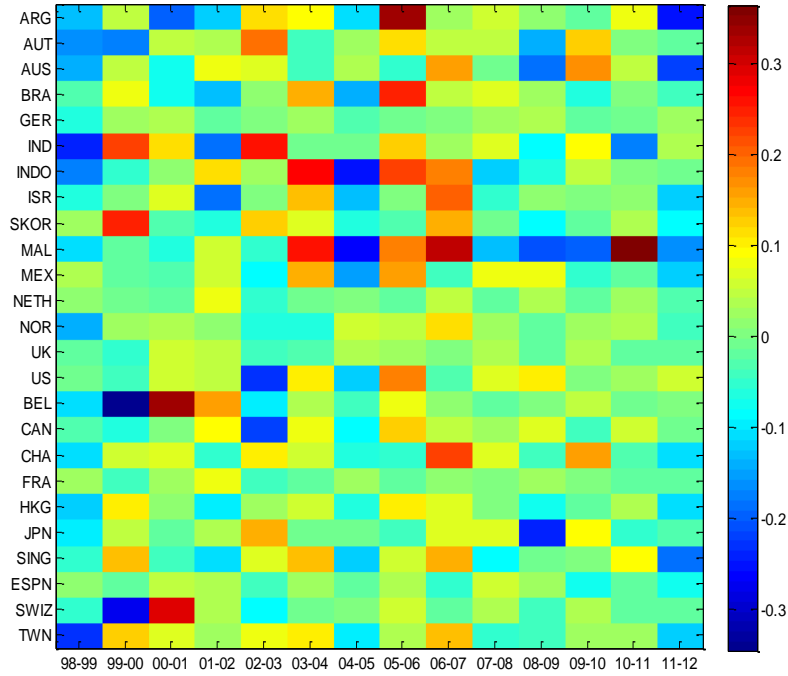

FIG. 2. Gray-scale representation of the changes in PC coefficients over time. Each index is denoted with a vertical line. The numbers along the horizontal axis indicate years. Dark or light blue areas indicate market recovery. Dark or light red areas indicate a crisis state. Yellow or green areas indicate a stable market.

V. Group Dynamics of the Markets

The loading plots of the first three PC coefficients calculated from correlations between principal components and returns (assets) are shown in Figs. 3-8. The aim of this projection is to visualize the relative two-dimensional positions of indices in space and to show the impact of the contribution (weight) of each index on each PC. The first PC implies the maximum variances of the data in the direction of the largest eigenvector, and the second PC indicates the second-largest variances in the direction of the second vector. Our experimental observation shows that the first three eigenvalues are greater than 1 over the study period. According to Kaiser's rule [27], eigenvalues greater than 1 are significant descriptors of data variance. As a result, we consider the first three PCs for further study. We focus on observing the group and its orientation in two-dimensional space due to crises, especially after 2005. In Fig. 3, we plot the first two PC coefficients of global financial indices in 1998 and 1999. We present the loading plot of global indices during the Russian crisis (in 1998) in Fig. 3(a) and that after the Russian crisis (in 1999) in Fig. 3(b). It is known that the coefficients of indices closest to one another and far from the origin of the plot are strongly correlated. Considering the nearest points in two-dimensional space, we classify the whole system into three regional groups: American, European, and Asian. The strong contribution of the European indices to the first and second PCs is clear over the observation periods. Moreover, all indices in the European zone contribute almost equally to the first and second PCs. However, the indices of the American and Asian groups do not show equal contributions to the first and second PCs (the points are scattered). The indices of Israel (8) and China (18) are not similar to those of any regional group. Additionally, the smallest contribution to the PC is observed in China's index, which is persistent throughout most of the period. The positions of indices in the loading plot change with time. We consider the center of mass of each group when observing its change in state. To identify the location of a group with respect to the principle axis, we measure $\cos(\theta)$, where $\theta$ is the angle between the line passing through the center of mass of each group and the principle axis.

This angle measures *how much a group deviates from the principle axis due to a crisis.* If the center of mass of a group is close to the principle axis, $\cos(\theta)$ approaches one. Since China and Israel do not fit into a group during the observation period, we do not consider these two countries when calculating the center of mass. We observe that the positions of the European cluster do not change significantly due to crisis or external effects. However, the American and Asian groups change positions due to crisis. When we compare the group dynamics during the Russian crisis to the dynamics after the crisis, we observe that the American group moves in the direction of the first principal axis with a change in angle from $\cos(x)=0.86$ to $\cos(x)=0.93$, and that the Asian group changes position in the direction of the second principal axis from $\cos(x)=0.70$ to $\cos(x)=0.54$. The European group changes position slightly in the direction of the second PC. Since the correlations between the first PC and the assets are strong during the crisis, a lower contribution of assets to the first PC implies recovery of the indices. After the Russian crisis, most of the indices in the American group move in the direction of the first PC toward a state similar to that before the crisis, with the exception of Argentina. During the crisis, the positions of all indices, except those of the Asian group, are nearer to one another in two-dimensional space than in normal periods. This implies an almost equal contribution of each asset to both PCs.

The loading plots for PC1-PC3 during the Russian crisis in 1998 are shown in Fig. 4(a). We observe a strong correlation among European countries. However, the American and Asian indices do not show distinct grouping during this time but instead are very scattered. After the Russian crisis in 1999, three distinct groups are formed (European group, Asian group, and American group), with Indonesia and China again presenting as outliers. The contribution of Asian assets to PC3 is smaller than those of other regional assets.

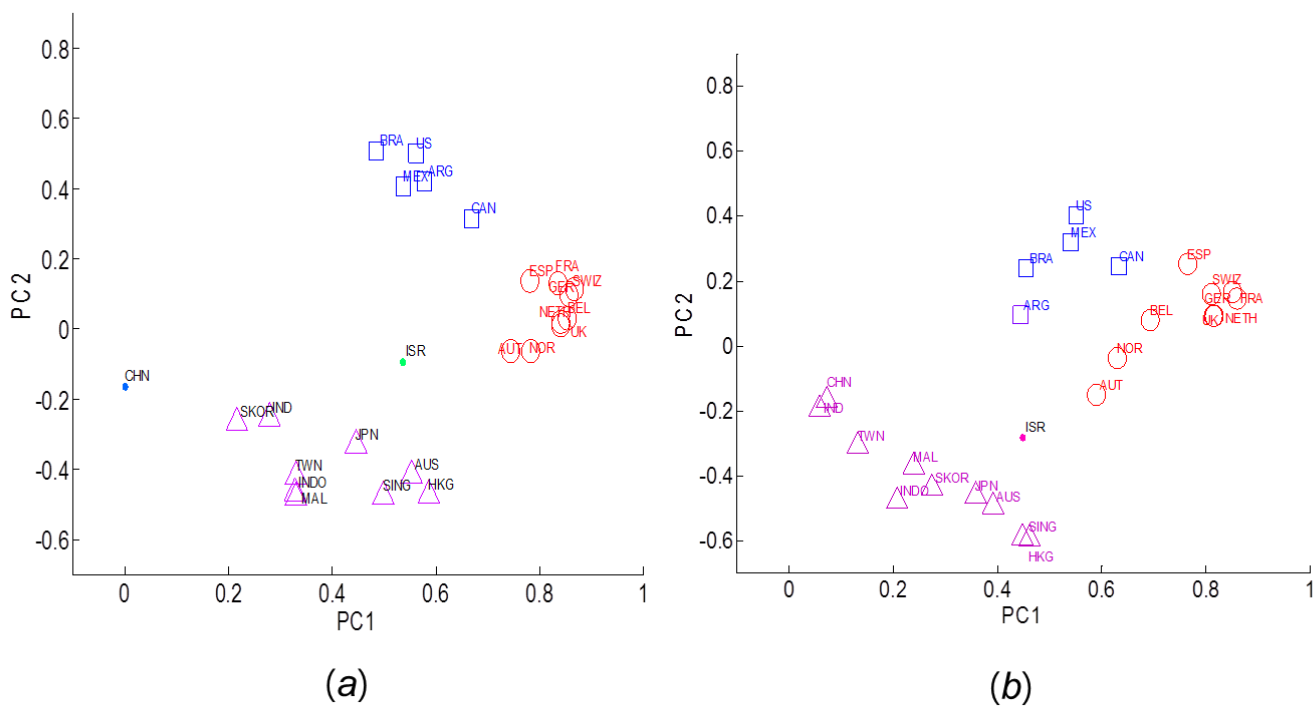

FIG. 3. Loading plots of PC1 and PC2 of global indices (a) during the Russian crisis in 1998 and (b) after the crisis in 1999. Three groups are clearly visible. The European group (red circle) is more robust than the other groups. Israel and China are outliers.

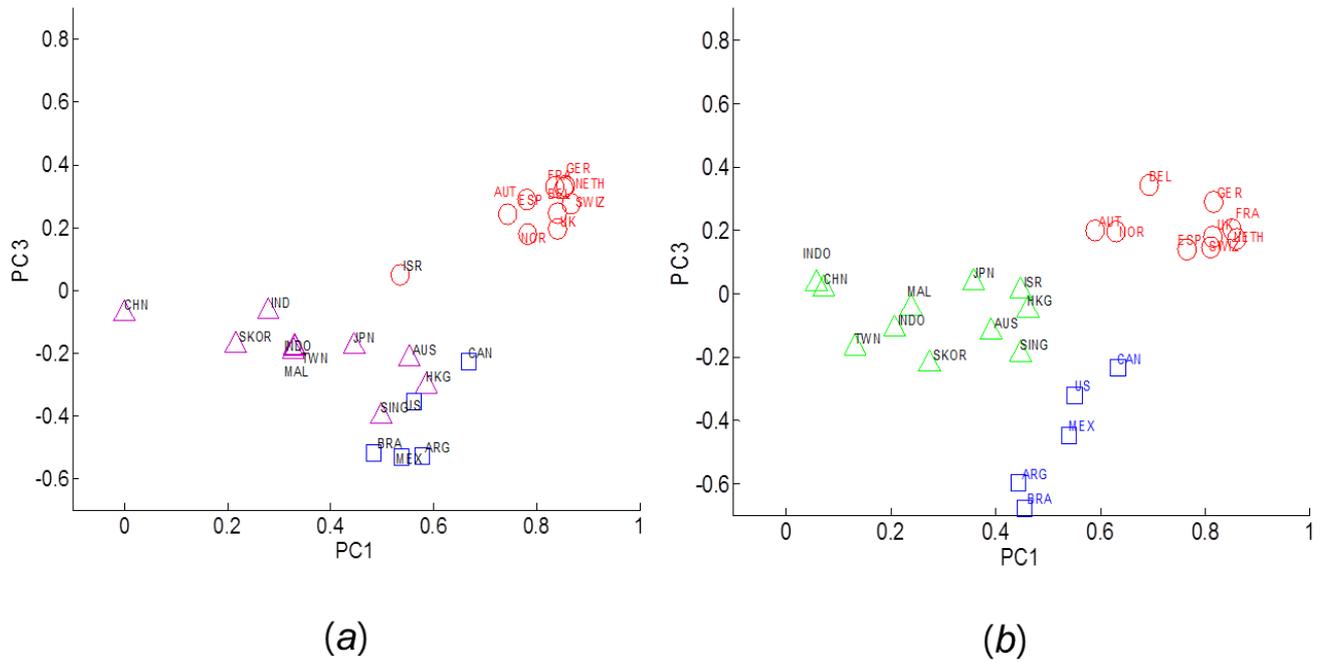

Fig. 4. Loading plots of PC1 and PC3 of global indices in (a) 1998 and (b) 1999. American and Asian indices are scattered. European indices are clearly grouped and strongly correlated.

We next analyze the group dynamics after the year 2005 and the spread of the mortgage crisis, global financial crisis and European sovereign crisis across the globe. We do not provide a detailed explanation of the group dynamics for the years 2000-2002. Since the effect of the dot-com bubble on the global market in these periods is not severe [28], the group dynamics do not show any significant effects or differences in this period.

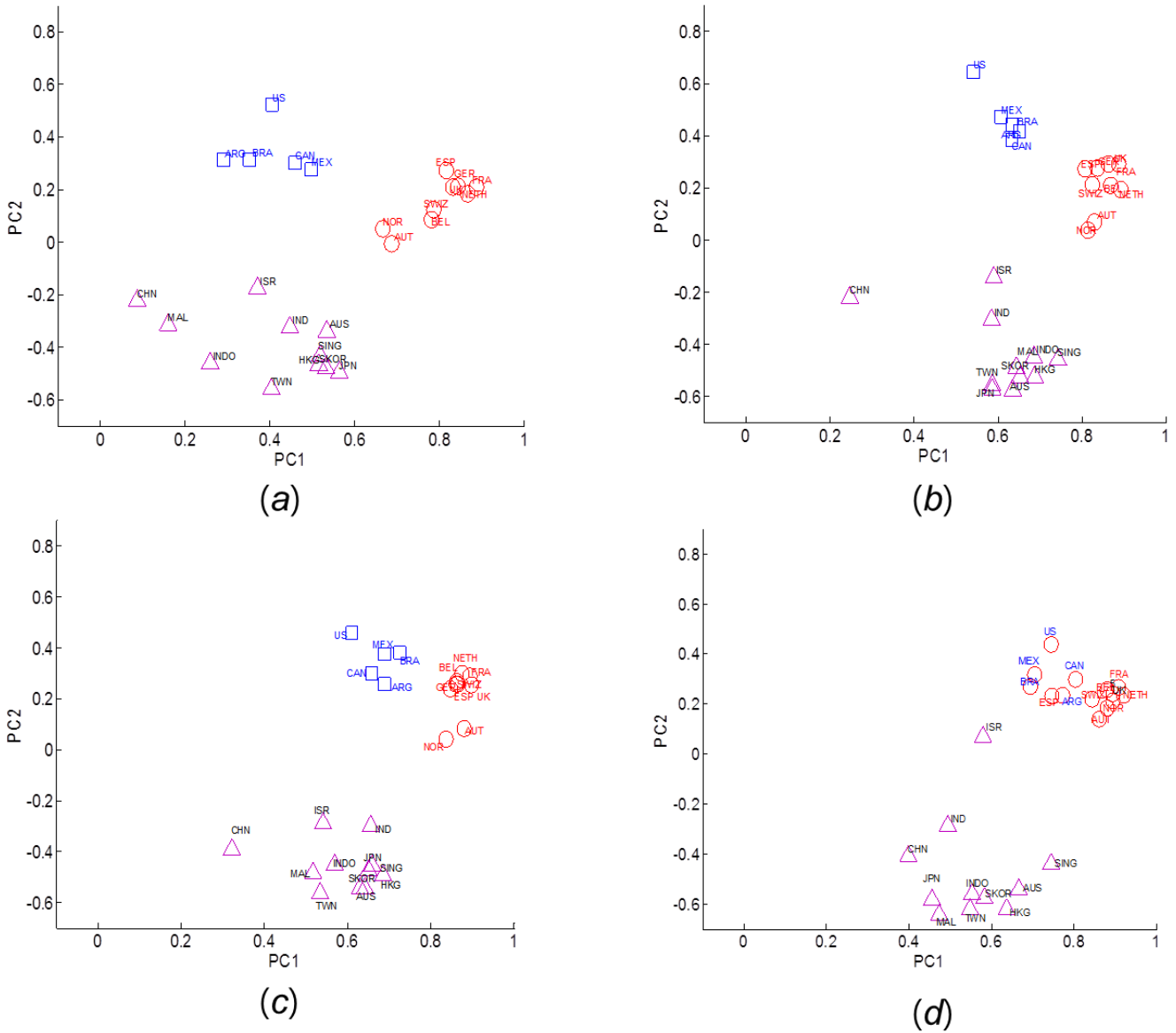

FIG. 5. Loading plots of PC2 against PC1 for global indices in the years (a) 2005, (b) 2007, (c) 2008, and (d) 2011. The indices of China, Israel, India, Austria, and Norway show high dispersion and are more similar to their regional indices during financial crises.

The year 2005 involved no special event in the global financial market. The loading plots of PC1-PC2 in this period show three groups, for which the indices of the American and Asian groups are marginally scattered. During the mortgage crisis in 2007, most of the indices in each group approach one another, which implies an almost equal contribution of each asset to the first and second PCs. The groups move in the direction of the largest PC due to the crisis. The angles of displacement of the groups relative to the centers of mass are given in Fig. 6. Significant change is observed in the American group during 2006 when the mortgage crisis originated, while the Asian group shows a sharp change in 2007. The angle ($\cos x$) shows sharp increases during the meltdown of the market and decreases during recovery periods. There is no significant change in the European indices. It is interesting that the indices of the regional groups are nearer to each other over the observation time window (comparing the figures from 1998 to 2012). This may be due to globalization. After the global financial crisis, the European and American indices approach each other and look like a single group (see Fig. 5(d)).

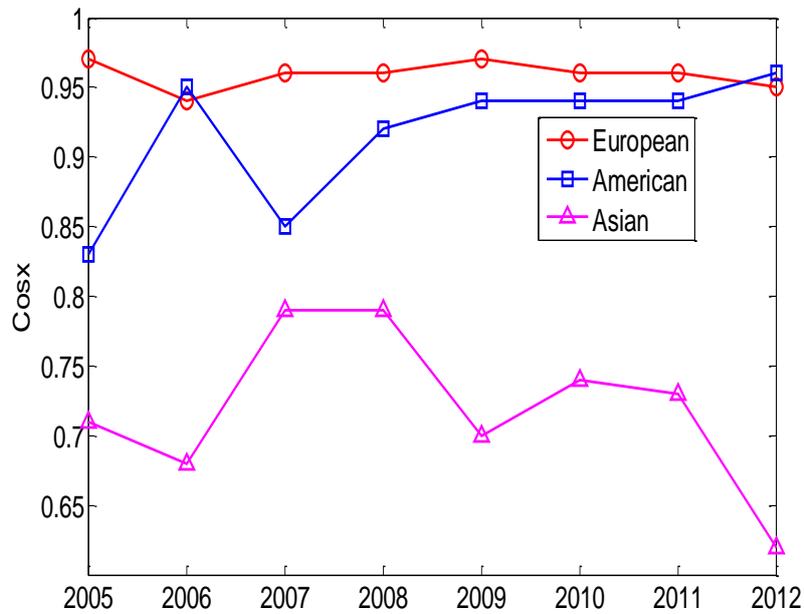

FIG. 6. The position of the group is measured by $\cos x$ as a function of year, where $x$ is the angle between the center of mass of a group and the first principal axis. We identify three distinct groups that correspond to the continental economic zones of Europe, America, and Asia.

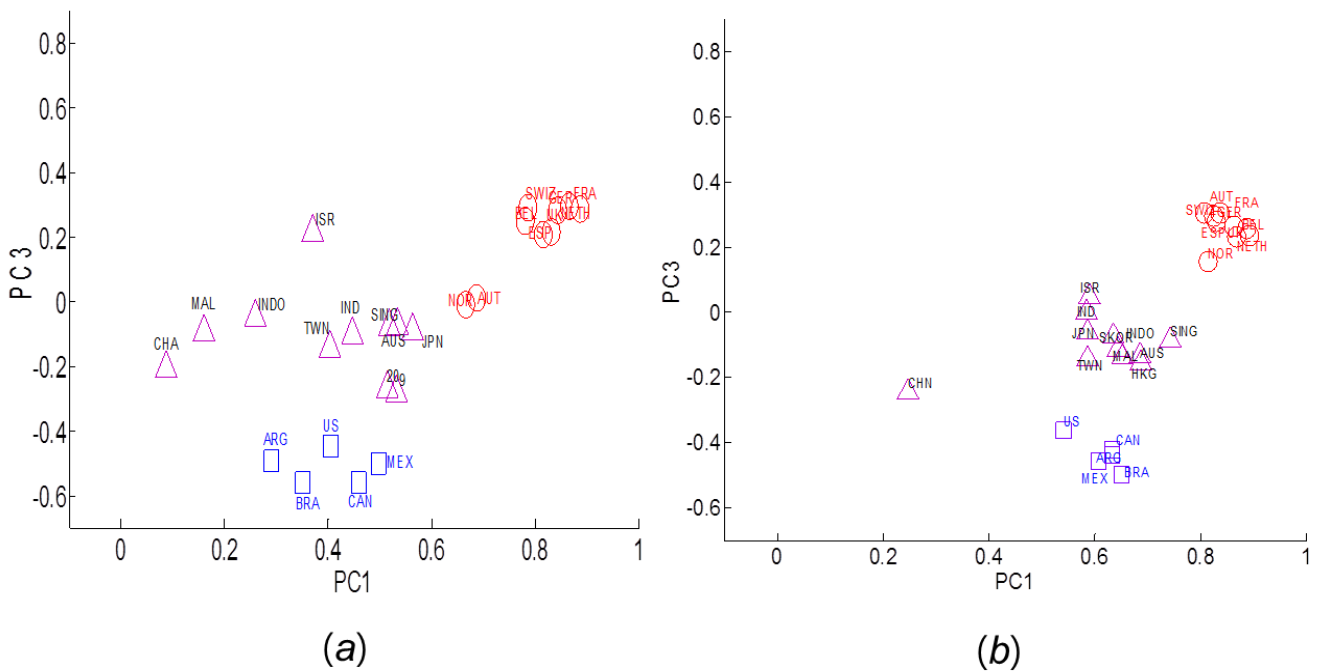

FIG. 7. Loading plot for PC1 and PC3 for (a) 2005 and (b) 2007. The indices are scattered in 2005, when there is no crisis. However, during the crisis in 2007, the indices are more tightly grouped.

The loading plots of PC1-PC3 are shown in Fig. 7. Three regional groups are obviously visible. As with PC1-PC2, large variances are also observed for the European group. This implies that the contributions of European indices to the third PC are more significant than the contributions of other groups. The center of mass of each group does not change much, as for PC1-PC2. When comparing a crisis year with a normal year, the indices of each group are more similar and create a strong group during the crisis. This implies that the contributions of group member assets to PC3 are similar. However, in most of the periods, the indices of China (18), Israel (8), Canada (17), Norway (13), Austria, and India are displaced from the groups. A plot of PC1-PC4 is presented in Fig. 8, showing that the European group is clearly isolated, while the American and Asian groups are distributed randomly in the two-dimensional projection. This indicates that only the eigenvalues greater than 1 (greater than PC4) explain significant variances in the original data matrices. Of course, there are some exceptions during the crisis period.

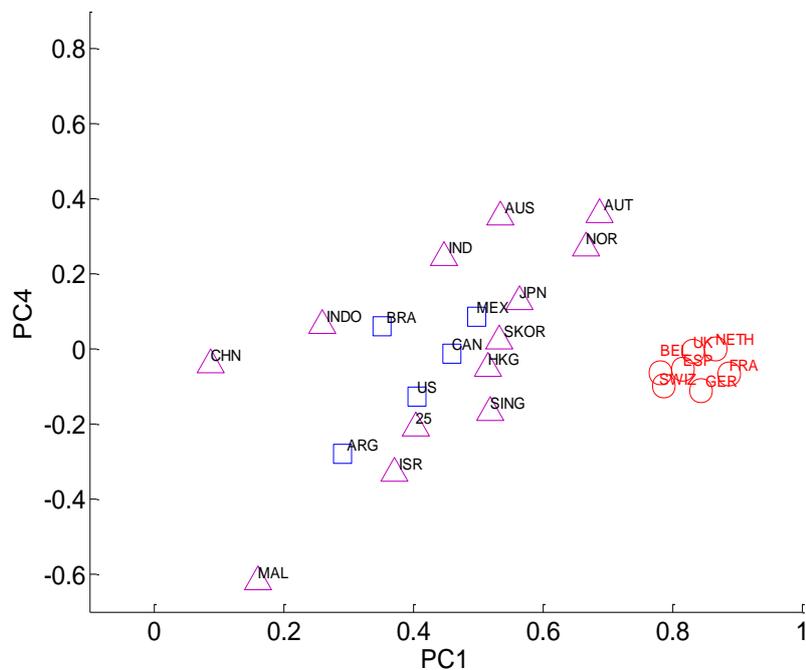

FIG.8. Loading plot of PC1-PC4 for the year 2005. The European group is completely isolated. The other indices are randomly distributed, and it is impossible to identify groups.

**VI. Conclusion**

The technique of principal component analysis is applied to the time series of global financial indices to investigate the correlation structure of different markets. The variances explained by PC are determined. The variances explained by the first PC increase with time and show a drastic change during the crisis. A sharp change in PC coefficient implies a transition of market state, a situation which occurs frequently in the American and Asian indices. However, the European indices remain stable over time. Using the first two PC coefficients, we identify three regional groups whose equity indices are closely related. The groups change position in two-dimensional spaces due to crises. We conclude that PCA is an effective tool to identify market state and to determine the subsets of global equity markets.

**ACKNOWLEDGEMENTS**

This research was supported by the Basic Science Research Program through the National Research Foundation of Korea (NRF), funded by the Ministry of Science, ICT and Future Planning (NRF-2014R1A2A1A11051982).